\documentclass[usegraphicx,useAMS]{mn2e}
\usepackage{graphics}
\usepackage{times}

\title[Radio observations of NGC 4254]
{Low radio frequency signatures of ram pressure stripping in Virgo spiral NGC 4254 \\}
\author[N. G. Kantharia et al.]{N. G. Kantharia$^1$,  A. Pramesh Rao$^2$,  S. K. Sirothia$^3$ \\
National Centre for Radio Astrophysics,  Tata Institute of Fundamental
Research, \\ Post Bag 3, Ganeshkhind, Pune-411007, India \\
$^1$ ngk@ncra.tifr.res.in \\
$^2$ pramesh@ncra.tifr.res.in \\
$^3$ sirothia@ncra.tifr.res.in \\
}

\begin{document}

\date{Accepted 2007 September 27. Received 2007 September 26; in original form 2007 May 8}

\maketitle

\begin{abstract}
We report the detection of extended low radio frequency continuum emission 
beyond the optical disk of the spiral galaxy NGC 4254
using the Giant Metrewave Radio Telescope.
NGC 4254, which has an almost face-on orientation, is located in the outskirts of the Virgo cluster.
Since such extended emission is uncommon in low inclination galaxies,
we believe it is a signature of magnetised plasma pushed out of the 
disk by ram pressure of the intracluster medium as NGC 4254 falls into
the Virgo cluster.  The detailed spectral index distribution across
NGC 4254 shows that the steepest spectrum 
$\alpha < -1$ ($S \propto \nu^{\alpha}$) arises in the gas beyond the
optical disk.  This lends support to the ram pressure scenario
by indicating that the extended emission is not from the disk gas 
but from matter which has been stripped by ram pressure.  The steeper spectrum of
the extended emission is reminiscent of haloes in edge-on galaxies.  The sharp fall in intensity
and enhanced polarization in the south of the galaxy, in addition to enhanced star formation
reported by others provide evidence towards the efficacy of ram pressure on this galaxy.  
HI 21cm observations show that the gas in the north lags in rotation and hence
is likely the atomic gas which is carried along with the wind.  
NGC 4254 is a particularly strong radio emitter with a power of
$7\times10^{22}$ Watts-Hz$^{-1}$ at 240 MHz.  We find that the integrated spectrum of the galaxy
flattens at lower frequencies 
and is well explained by an injection spectrum with $\alpha_0=-0.45\pm0.12$.  
We end by comparing published simulation results with our data and
conclude that ram pressure stripping is likely to be a significant contributor  to 
evolution of galaxies residing in X-ray poor groups and cluster outskirts. 
\end{abstract}

\section{Introduction}
Galaxy mergers (Toomre \nocite{toomre} 1977), galaxy harassment (Moore et al. \nocite{moore} 1996), 
ram pressure (Gunn \& Gott \nocite{gunn} 1972), viscous stripping (Nulsen 1982 \nocite{nulsen})
and dynamical friction (Lecar \nocite{lecar} 1975, Chandrasekhar \nocite{chandrasekhar} 1943) 
are believed to play a role in the evolution of galaxies
in a cluster enviroment and have explained numerous observational results.  Different physical
mechanisms dominate in different environments.  While ram pressure effects are
believed to be dominant in the dense inner regions of clusters, tidal interactions
are believed to dominate the evolution in the cluster outskirts and in groups of galaxies.
However, this picture has been undergoing a gradual paradigm shift as demonstrated
by observations, especially at radio frequencies, and simulations
of groups and outskirts of clusters (e.g. Virgo) conducted by
various authors. 

The Virgo cluster galaxies have
been extensively studied in radio and other wavebands e.g. in HI 21cm
by Davies, et al. \nocite{davies} (2004), Cayatte et al. \nocite{cayatte} (1990), 
Warmels \nocite{warmels} (1988), Huchtmeier \& Richter \nocite{hucht} (1986) and in
radio continuum by Vollmer et al. \nocite{vollmer2} (2004).  Chung et al. \nocite{chung} (2007)
have recently reported HI tails for several Virgo galaxies which they conclude is ram pressure
stripped gas from the galaxy.  Vollmer et al. \nocite{vollmer3} (2007)
have studied the polarisation properties of several Virgo spirals and 
inferred that the interaction of the galaxy with the intracluster medium (ICM)
has resulted in peculiar polarisation properties of the cluster spirals. 
Wezgoweic et al. \nocite{wezgoweic} (2007) 
find distorted magnetic fields in several Virgo cluster members which
they attribute to ram pressure of the ICM. 

NGC 4254 is an interesting almost face-on SA(s)c spiral, located in the outskirts of the Virgo cluster,
with one dominant spiral arm.  This galaxy 
has been studied extensively in the HI line (Cayatte et al. \nocite{cayatte} 
1990, \nocite{cayatte1} 1994,  Phookun et al. \nocite{phookun} 1993),
in high frequency radio continuum and polarization  (Urbanik et al. 
\nocite{urbanik} 1986, Soida et al.  \nocite{soida} 1996, Urbanik \nocite{urbanik1} 2004
Chyzy et al. \nocite{chyzy} 2006), 
in X-rays (Chyzy et al. 2006 \nocite{chyzy}, Soria \& Wong 2006 \nocite{soria})
and via simulations involving tidal interaction and ram pressure
stripping (Vollmer et al 2005 \nocite{vollmer}).
NGC 4254 is both optically and radio bright and lies about $4^{\circ}$ ($\sim 1.2$ Mpc
using a distance of 17 Mpc to Virgo cluster) to the north-west of M87.
Phookun et al. \nocite{phookun}(1993), from their deep HI observations, separate
the disk and non-disk emission from the galaxy and find that the latter forms a tail of clouds that
extend to $\sim 11'$ from the galaxy.
They conclude that there is an infall of a disintegrating cloud of gas onto NGC 4254
and that the one-armed structure is a result of the tidal
interaction with the infalling gas.
Vollmer et al \nocite{vollmer}(2005) have modelled NGC 4254 after
including a tidal encounter and ram pressure stripping.  They find that the one-armed
structure can be explained by a rapid and close encounter $\sim 2.8\times10^8$ years ago  
with another massive galaxy ($\sim 10^{11} M_\odot$).  They note that most of the observed
HI morphology is reproduced by including ongoing ram pressure effects in the
simulation and only two main HI features remain unexplained namely   
1) the shape of the extended tail of HI emission  in the north-west,
2) a low surface density HI blob observed to the south of the main disk.
Davies et al.  \nocite{davies} (2004) and Minchin et al. \nocite{minchin} (2005) have 
reported the detection of a massive ($\sim 10^{11} M_{\odot}$)
HI cloud VIRGOHI 21 in the Virgo cluster.  This cloud lies to the north of NGC 4254 and
connects both spatially and kinematically to NGC 4254 
(Minchin et al. \nocite{minchin1} (2005).  Recently Haynes et al.
\nocite{haynes} (2007) have presented the HI map of this entire region made using the Arecibo
telescope.  They detect a long HI tail which starts from NGC 4254, includes 
VirgoHI 21 and extends northwards to a total distance of 250 kpc.  
Earlier observations had only detected parts of this long HI tail. 
Haynes et al. \nocite{haynes} (2007) attribute this long HI tail 
to galaxy harrassment (Moore et al. \nocite{moore} 1996) which the galaxy is undergoing as
it enters the Virgo cluster.

Soida et al. \nocite{soida} (1996), from their
radio continuum observations near 5 and 10 GHz, have reported enhanced polarisation
in the southern ridge which they attribute to interaction of the disk gas with the ICM. 
Chyzy et al. \nocite{chyzy} \nocite{chyzy1} (2006,2007) detect an extended polarized envelope at 1.4 GHz. 
Excess blue emission is observed along the southern ridge (Knapen et al. \nocite{knapen} 2004)
which indicates the presence of a large population of young blue stars (Wray \nocite{wray} 1988).  
The trigger for the enhanced star formation in this region
is possibly compression by the ram pressure of the ICM.  
From high resolution CO data, Sofue et al. \nocite{sofue} (2003) note that the  
inner spiral arms of NGC 4254 are asymmetric.  They reason that the ram pressure of the ICM
has distorted the inter-arm low density regions leading to the
asymmetric spiral arms. 
This galaxy has also been observed in the near and far ultraviolet by
GALEX (Gil de Paz et al. \nocite{gil} 2007).  The UV
morphology of the galaxy is similar to the DSS optical with several young
star forming regions visible in the GALEX images.

In light of all these interesting results, we present new low
frequency radio continuum and HI 21cm observations of NGC 4254 using the Giant Metrewave
Radio Telescope (GMRT) near Pune in India.  The continuum observations were conducted 
at four frequencies between 240 and 1280 MHz.  We detected extended 
continuum emission surrounding the optical disk of NGC 4254. 
We discuss the integrated spectrum of the galaxy and
the environmental effects on this galaxy located at the periphery of the cluster.
We adopt a distance of 17 Mpc to the Virgo Cluster following Vollmer et al. 
\nocite{vollmer} (2005).  In Table \ref{tab1}, we summarize the physical properties
of NGC 4254 and the empirical data from wavebands ranging from
X-rays to radio. 

\begin{table}
\caption{Properties of NGC 4254}
\begin{tabular}{lcc}
\hline\hline
{\bf Parameter}  & {\bf Value/Property} &  {\bf Reference} \\
$\alpha(2000)$  & $\rm 12^h 18^m49.6^s$  &  RC3  \\
$\delta(2000)$  & $14^\circ24'59''$ &  RC3 \\
Morphological type  & Sc &  RC3 \\
Optical diameter D$_{25}$ & $5.4'$ &  RC3 \\
B$^0_T$  & 10.44 &  RC3 \\
Heliocentric velocity & $2407\pm3$ kms$^{-1}$ &  RC3 \\
Distance & $ 17$ Mpc & Vollmer et al. (2005) \\
Distance to cluster centre & $\sim 1.2$ Mpc &   \\
\hline
\multicolumn{3}{l}{\bf From HI} \\
Morphology & distorted & Cayatte et al. (1990) \\
V (rotation)  & 150 kms$^{-1}$ & Phookun et al. (1993) \\
Inclination angle & $42^{\circ}$ & Phookun et al. (1993) \\
Non-disk emission & $2.3\times10^8 M_\odot$ & Phookun et al. (1993) \\
HI deficiency & $0.17\pm0.2$ & Cayatte et al. (1994) \\
ICM ram pressure &$0.9\times10^{-12}$ dyn-cm$^{-2}$  &  Cayatte et al. (1994)\\
Grav pull & $1.6\times10^{-12}$ dyn-cm$^{-2}$  & Cayatte et al. (1994)\\
Galaxy Harrassment & 250 kpc tail & Haynes et al. (2007) \\
\multicolumn{3}{l}{\bf From radio continuum}  \\
Radio spectrum (0.4-10 GHz) & $-0.83\pm0.04 $ & Soida et al. (1996) \\
Polarised emission & excess in south & Soida et al. (1996) \\
Polarised envelope & yes & Chyzy et al. (2007) \\
\multicolumn{3}{l}{\bf From Optical (DSS),   NIR (2MASS), NUV,FUV (GALEX) }   \\
Structure & One-arm morphology & ...\\
Star formation & Vigorous & ...\\
Blue stars & Forms a ridge in south & Wray (1988) \\
B Band & Excess in south & Knapen et al. (2004) \\
\multicolumn{3}{l}{\bf From CO}  \\
Inner arms & asymmetric & Sofue et al. (2003)\\
Possible Reason & ICM ram pressure & Sofue et al. (2003)\\
\multicolumn{3}{l}{\bf From X-ray}  \\
Galaxy & Intense emitter & Fabbiano et al. (1992) \\
Discrete source & ULX in south & Soria \& Wong (2006)\\
\multicolumn{3}{l}{\bf From simulations}  \\
One arm & Tidal  & Vollmer et al. (2005) \\
Obs. HI structure & tidal+ram press & Vollmer et al. (2005) \\
Stripping angle & $70^{\circ}$ & Vollmer et al. (2005) \\
\hline\hline
\end{tabular}
\label{tab1}
\end{table}

\section{Observations and Results}
NGC 4254 was observed at 240 MHz, 325 MHz, 610 MHz and 1280 MHz
in the radio continuum and in the 21 cm emission line of HI  
using the GMRT (Swarup et al. \nocite{swarup}1991,  
Ananthakrishnan \& Rao \nocite{ananth} 2002).
GMRT consists of thirty 45-m diameter parabolic dish antennas spread 
over a 25 km region and operates at five frequency bands below 1.5 GHz.
Details of our observations are listed in Table \ref{tab2}.

\begin{table*}
\caption{Details of radio observations.  The integrated flux density of NGC 4254
is given in the last row.  }
\begin{tabular}{lccccc}
\hline\hline
Obs Freq (MHz)  &  240 & 325  & 610   & 1280  & 21 cm HI\\
Obs Dates &  20/02/06 & 24/04/04 & 20/02/06 & 23/10/00 & 16/11/00  \\
Working antennas &  23   & 26   &  23    &  18  & 22  \\
Bandwidth(MHz)  & 5 & 9   &  13   &  12 & 0.0625$^1$ \\
{\bf Robust=0}  &&&    &   & \\
Synthesized Beam & $17.7''\times11.4''$ & $27''\times10.3''$ & $9.9''\times 5.3''$  
& $23'' \times 20''$$^2$ & $7.5'' \times 4.9''$      \\
Position angle & $-81^{\circ}$  & $-47^{\circ}$ & $-87^{\circ}$ & $-44^{\circ}$ & $31^{\circ}$  \\
RMS noise (mJy/beam) &  1.1  &  1.0 &  0.2  & 0.1  & 2.3$^3$ \\
{\bf Robust=5} && &    &   & \\
Synthesized Beam & $24''\times21''$ & $34''\times20''$ & $17''\times 16''$  & $19'' \times 17''$ &
   -      \\
Position angle & $-31^{\circ}$  & $-6^{\circ}$ & $24^{\circ}$ & $54^{\circ}$ & -  \\
RMS noise (mJy/beam) &  0.9  &  2 &  0.2  & 0.15  & - \\
\hline
S (Jy)  & 1.96 & 1.6   & 1.08  & 0.377  &  - \\
\hline\hline
\end{tabular}

$^1$ 128 channels over the 8 MHz bandwidth resulted in this channel width 
(i.e. $12.5$ kms$^{-1}$).

$^2$ using data on baselines upto $\rm 14 k\lambda$.

$^3$ rms is for a channel of width 12.5 kms$^{-1}$.

\label{tab2}
\end{table*}

\subsection{Radio continuum}

The visibility data obtained at GMRT in `lta' format were converted to FITS and 
imported to AIPS\footnote{AIPS is distributed by NRAO
which is a facility of the NSF operated under cooperative agreement by Associated Universities, Inc. }.  
The analysis was carried out using standard tasks in AIPS. 
Since the field-of-view at 610, 325 and 240 MHz is large, the images were generated
using multiple facets.  
The final images were generated using robust weighting (Briggs \nocite{briggs} 1995) of
the visibilities.  We made the images using 
robust=0 (between pure uniform weighting i.e. robust$=-5$ and pure
natural weighting i.e. robust$=5$) and robust=5.   

\begin{figure*}
\caption{ Maps of NGC 4254 made using Briggs robust=0.  The contours are plotted
at 3, 6, 12, 24, 48 and 96 times the rms noise $\sigma$.
The crosses mark the position of the optical centre of NGC 4254 and the three background point
sources in the field.
(a) Top left panel.  The image at 240 MHz with a beamsize $17.7'' \times  11.4''$, 
PA$= -81^{\circ}$
The rms noise is $\sigma = 1.1$ mJy/beam.  The highest contour is $35\sigma$.
(b) Top right panel. The image at 325 MHz.  The angular resolution is
 $ 27'' \times 10.3''$, PA$= -46.8^{\circ}$.  The rms noise is
$\sigma = 1$ mJy/beam and the highest contour is $35\sigma$.
(c) Bottom left panel. The image at 610 MHz has a resolution  of
$9.9''\times 5.3''$, PA$= -87^{\circ}$.  The rms noise is 
$\sigma=0.2$ mJy/beam and the highest contour is $24\sigma$.
(d) Bottom right panel. Image at 1280 MHz, obtained with a uvtaper of $14 k\lambda$ 
along both axes (u,v) and convolved to an angular resolution of 
$23'' \times 20''$, PA$= -44^{\circ}$ to highlight 
the extended emission in the north-west.
The rms noise is $\sigma = 0.1$ mJy/beam and the highest contour is 140$\sigma$. }
\label{fig1}
\end{figure*}

In Fig. \ref{fig1}, we show the images made using robust=0.
A sharp cutoff to the radio emission is observed to the south/south-east (see the 325 and
1280 MHz images in Fig. \ref{fig1}).
A radio continuum minimum is detected in the 
north-west at all frequencies although the extent and shape varies.  
Part of this difference is likely due to the different spatial frequency coverage
of the multi-frequency observations.  This minimum is accentuated in the naturally
weighted maps with it being most prominent at 240 MHz (see Fig. \ref{fig3}b).   
The 1280 MHz image (robust=0) was made using a maximum baseline of 14 k$\lambda$ 
to highlight the extended features. 
In Fig. \ref{fig2}, the 610 MHz image has been superposed on the DSS optical image
to show the envelope of radio emission around the optical disk. 
We note that our images at 240 MHz and 610 MHz made using robust=5
are by far the most sensitive low frequency images of this galaxy with rms noise
of 0.9 mJy/beam and 0.2 mJy/beam respectively, for the beamsizes listed in Table \ref{tab2}.  
The signal-to-noise ratio on the robust=5 image is higher
than on the robust=0 image by a factor of about 3.
In Fig. \ref{fig3}, the robust=0 and robust=5 images at 240 MHz are shown for comparison.
The correlation between the images is good as seen from the radio emission which extends outwards
of the optical disk in the north-west, south-east and south-west and is present in both the maps. 
For rest of the discussion, we use the naturally weighted images at 240 and 610 MHz 
unless stated.  

The integrated flux density of NGC 4254 at different frequencies are 
listed in Table \ref{tab2}.   
Visibility data corresponding to the three confusing sources (marked in 
Fig. \ref{fig3}) which are engulfed by the radio emission from NGC 4254 were removed 
before generating the images used for constructing the spectral index maps. 

\begin{figure}
\caption{The 610 MHz Briggs robust=0 map (in contours) is superposed 
on the DSS grey scale map.  Note the radio envelope around the optical disk.  
The crosses mark the position of the optical centre of the galaxy and the three background point sources.}
\label{fig2}
\end{figure}

\subsection{21 cm HI}

An 8 MHz bandwidth spread over 128 channels and centred at 2407 kms$^{-1}$
resulted in a channel width of 62.5 KHz $\approx$ 12.5 kms$^{-1}$ for our HI observations.  
3C147 and 1120+143 were used as the flux and phase calibrators respectively.
The HI data were analysed in AIPS++ ver 1.9 (build 1460).
The continuum was estimated by fitting a first order baseline to the
line-free channels and subtracted from the data.  A spectral line 
cube was then generated from this database. 
The moment maps shown in Fig. \ref{fig4} were obtained using a rms noise cutoff of
$1.5 \sigma$ in the channel maps.  The full resolution maps had an angular resolution of
about $7.5''\times4.9''$ with a position angle of $31^{\circ}$.

The HI distribution (Fig. \ref{fig4}) across NGC 4254 shows emission detected
from the disk but little HI is coincident with the
radio continuum extensions in the south-east, the north-west and the 
south-west.  The lowest column density is $3.4\times 10^{19}$ cm$^{-2}$.
The velocity field (Fig. \ref{fig4}b) shows that the north-east is the
receding side of the disk.  A slight lopsidedness is visible
in the velocity field with the north-east showing lower rotation 
($ < 100$ kms$^{-1}$) velocities compared to the south-west ($ < -100$ kms$^{-1}$)
wrt to the systemic velocity ($2407$ kms$^{-1}$) of the galaxy.
Fig. \ref{fig4}c shows the velocity dispersion across the galaxy.  Widest spectral lines
are seen near the centre whereas rest of the disk shows line widths of $10-15$ kms$^{-1}$.
The disk has a smooth boundary to the south, south-east whereas the northern boundary 
is jagged as if this part of HI is being `blown' off.  Interestingly, this is 
also the region where Phookun et al. \nocite{phookun} (1993) detect the largest
non-disk HI cloud.  
We do not detect the non-disk HI clouds (Phookun et al. 1993) or the long HI tail 
(Haynes et al. \nocite{haynes} 2007) associated with NGC 4254 likely due to 
lack of short spatial frequencies resulting in lower sensitivity to such extended faint features.  

\begin{figure*}
\includegraphics[width=8cm]{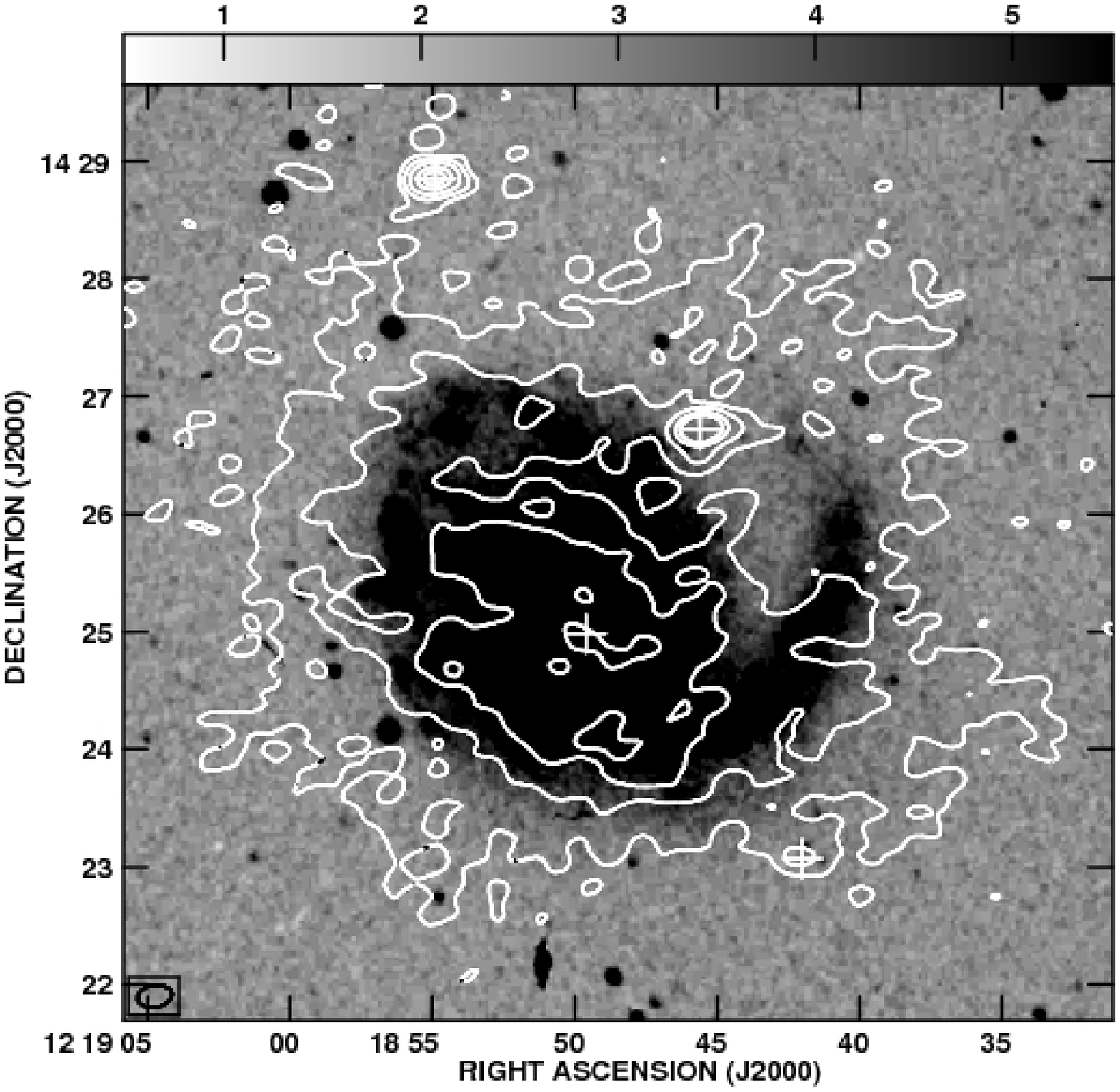}
\includegraphics[width=8cm]{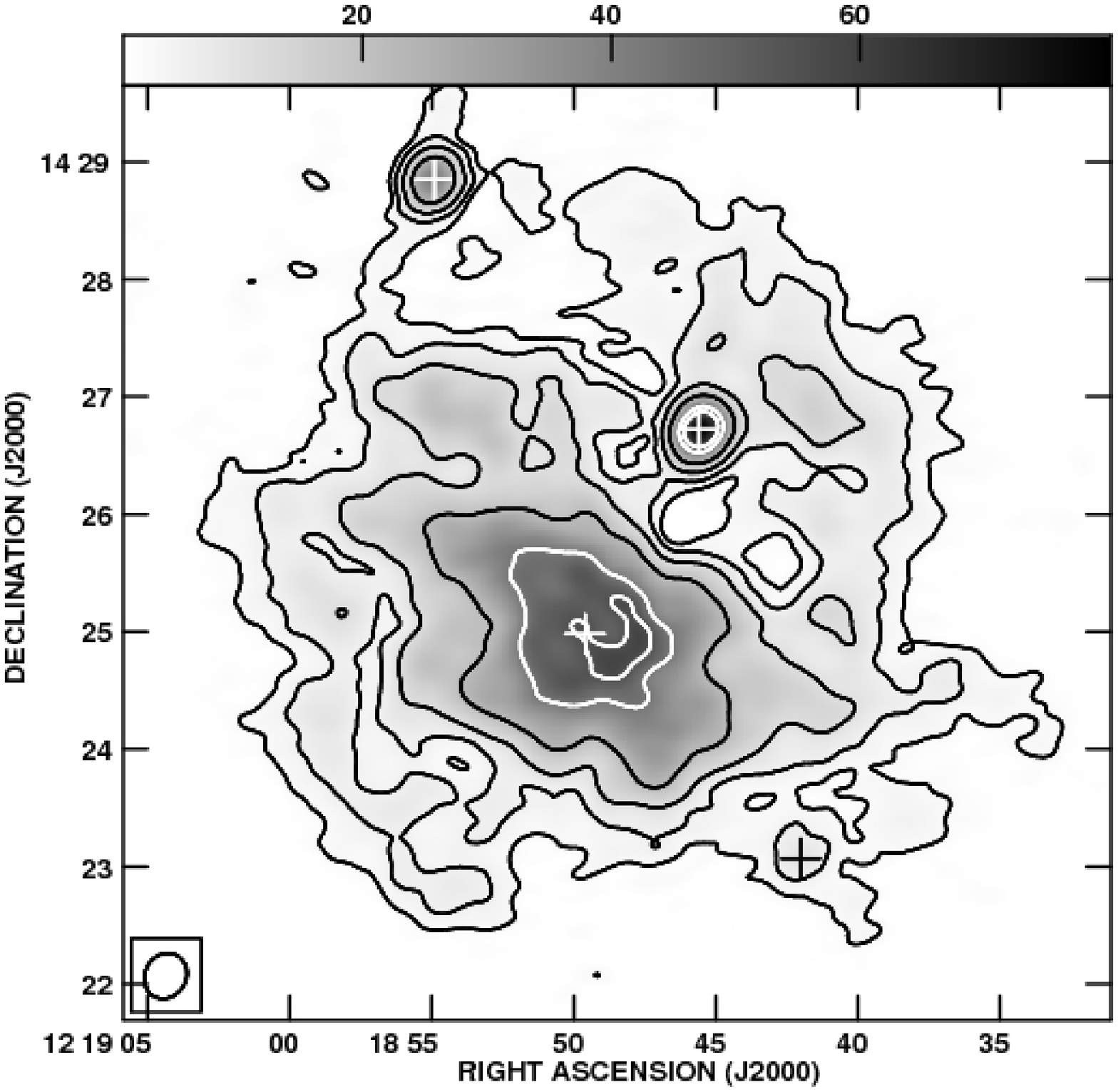}
\caption{240 MHz image.
Notice the envelope of radio continuum emission
at 240 MHz which surrounds the DSS map in the left panel.  The
contours are plotted at 3, 6, 12, 24, 48 times the rms noise ($\sigma$).
(a) The 240 MHz map made with robust=0 superposed on the DSS optical
map.  Angular resolution is
$17.7'' \times  11.4''$, PA$= -81^{\circ}$.  The rms noise is
1.1 mJy/beam.  The highest contour is $35\sigma$.
(b) The naturally weighted (robust=5) 240 MHz map.  The angular 
resolution is $24'' \times 21 ''$, 
PA$= -30^{\circ}$.  The rms noise is 0.9 mJy/beam.  The
highest contour is $72\sigma$.  The grey scale runs from 0.9 to 80 mJy/beam.}
\label{fig3}
\end{figure*}

\begin{figure}
\caption{ The HI moment maps are superposed on the 240 MHz
grey scale image. (a) Top panel. HI Moment 0 map.
The lowest contour level is $3.4\times 10^{19}$ cm$^{-2}$. 
Contours levels are $(20.4 + 2.43n)\times 10^{19}$ cm$^{-2}$ 
where n runs from -7 to 7.  (b)Middle panel.  HI Moment 1 map.  The systemic velocity
of the galaxy is 2407 kms$^{-1}$.  The solid lines indicate recession
while the dashed lines identifies the approaching side. (c) Bottom
panel.  HI Moment 2 map.  }
\label{fig4}
\end{figure}

\section{The Radio Morphology} 
We enumerate below the interesting morphological features
detected in the radio continuum emission (Figs. \ref{fig1}, \ref{fig2}, \ref{fig3}) and 
the HI (Fig. \ref{fig4}) from NGC 4254. 
\begin{enumerate}
\item Radio continuum emission and HI are detected from the optical disk.
\item Bright abrupt cutoff in the radio emission to the south of the galaxy which outlines the
optical disk.  
\item Low surface brightness emission enveloping the disk emission
is detected at all frequencies.   These features
are prominent at 240 MHz (see Fig. \ref{fig3}).  
Most notable features are the extension to the north-west, 
south-east and the finger-like extensions in the south-west which
is likely the gas blown out by the ICM pressure.
No R band or H$\alpha$ emission is detected from these regions.
\item A minimum in the radio continuum is detected in the north-west at all
the frequencies.  This minimum is most extended and prominent at 240 MHz.
Chyzy et al.  \nocite{chyzy} (2006,2007) report low polarization in the region which appears to be
partially coincident with the minimum at 240 MHz.  Thus the minimum could likely be due
to a decrease in the magnetic field in that region.
\item  About 15\% of the total emission at 240 MHz arises in the emission which surrounds
the optical disk. 
\item The HI disk is extended in the north-east (see Fig. \ref{fig4}).  
\item The HI velocity field is lopsided.  Lower recessional velocities are recorded
in the north-east and the same region also shows lower column densities.  
Moreover, the velocity field in the north appears to lag.
\item The HI has a jagged border in the north, north-west  (see Fig. \ref{fig4})
suggesting that the atomic gas is being progressively `blown' off. 
A smooth boundary is seen in the south, south-east. 
Moreover the HI shows a vertical edge in the north-east and the
west as if gas has been stripped off.   
\item HI is not found to be coincident with the radio continuum envelope 
around NGC 4254 (see Fig. \ref{fig4}).
\end{enumerate}

\begin{figure}
\includegraphics[width=6cm,angle=-90]{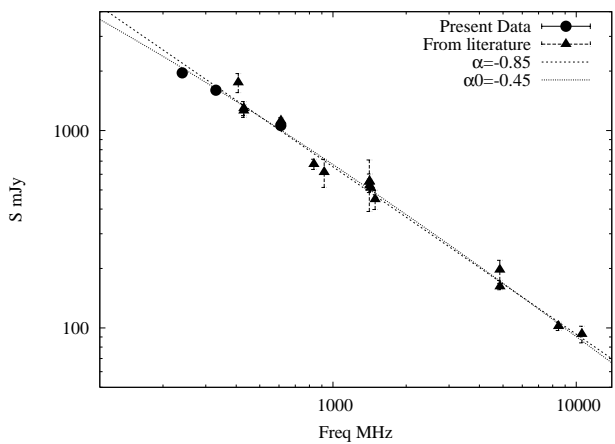}
\caption{Integrated spectrum of NGC 4254 including the
data from Soida et al. (1996) \& Chyzy et al. (2007).  The best single power law fit to
the data has an index $\alpha=-0.848$ and is shown by the dashed line.
The dotted line shows the best fit to the data if electron energy losses
are included.  It has an injection spectrum of $-0.45$ and has
lower $\chi^2$ than the single power law fit.  Data at frequencies $\le 100$ MHz are required to
confirm the break in the spectrum. }
\label{fig5}
\end{figure}

We compare the radio continuum envelope to the HI low resolution map in
Phookun et al. \nocite{phookun}(1993) who detected 
disk and non-disk HI components.  Phookun et al. \nocite{phookun} (1993)
inferred that the non-disk HI clouds around NGC 4254 were an infalling population.  
A similar conclusion can be drawn from the HI velocity field of
the non-disk clouds which are likely to be located between us and the galaxy. 
The HI blob (Fig 5 in Phookun
et al. 1993) to the south of NGC 4254 is coincident with part of southern extension we
detect at 240 MHz.  Moreover, some of the northern HI clouds are 
coincident with the northern radio continuum extension (see Fig \ref{fig3}).  
However, it is not clear if these are physically associated or
coincident in projection.  The HI velocity field is distorted in the north of the galaxy 
(Fig 4 (b) in Phookun et al. \nocite{phookun} 1993).  

\section{The Radio Spectrum}
We estimate a global spectral index between 240 and 610 MHz of $-0.65$. 
However, fitting a single power law using
the least squares method to all the available data ranging from 408 MHz to 
10 GHz (Soida et al. \nocite{soida} 1996, Chyzy et al. \nocite{chyzy1} 2007) and 
including our data (excluding the 1280 MHz point) resulted in the best fit spectrum
with an index of $-0.848 \pm 0.016$ (see Fig \ref{fig5}). This is in 
good agreement with the value ($-0.83\pm0.04$) that Soida et al. \nocite{soida} (1996) 
quote.  However  $\alpha=-0.65$ obtained from our low frequency data
suggests that the spectrum might be flattening at the lower frequencies.  
Since the galaxy is face-on, absorption by foreground thermal gas is expected to be 
negligible.   Thus, the break could possibly be due to propagation effects of
relativistic electrons and subsequent energy losses as explained by
Pohl et al. \nocite{pohl} (1991) \& Hummel \nocite{hummel} (1991). 

To check this, we fitted the observed integrated spectrum of NGC 4254 using the heuristic model given by 
Hummel \nocite{hummel} (1991) which describes the radio spectrum as 
\begin{equation}
S(\nu)= K \frac{\nu^{\alpha_0}}{1+ (\nu/\nu_b)^{1/2}}  
\end{equation}
The spectral index of the injection spectrum, $\alpha_0$ and 
the break frequency, $\nu_b$ are estimated
by fitting the observed spectrum with the above function.   
The best fit which has $\alpha_0=-0.45\pm 0.12$ is shown in Fig. \ref{fig5}.
This spectrum gives a lower $\chi^2$ compared to the single power law
fit (also in Fig. \ref{fig5}) and hence, we believe, is a better fit to the 
observed data between 240 MHz and 10 GHz.  However the fit fails to constrain the break frequency 
since we do not have data points below 240 MHz. 
From the above, it is reasonable to conclude that the integrated spectrum of NGC 4254
has a break at lower frequencies  which is typical of spiral galaxies
(Pohl et al. \nocite{pohl} 1991,  Hummel \nocite{hummel} 1991).
We also note that lower frequency data ($\nu\le100$ MHz) are required to confirm the break.

\begin{figure*}
\includegraphics[width=8cm]{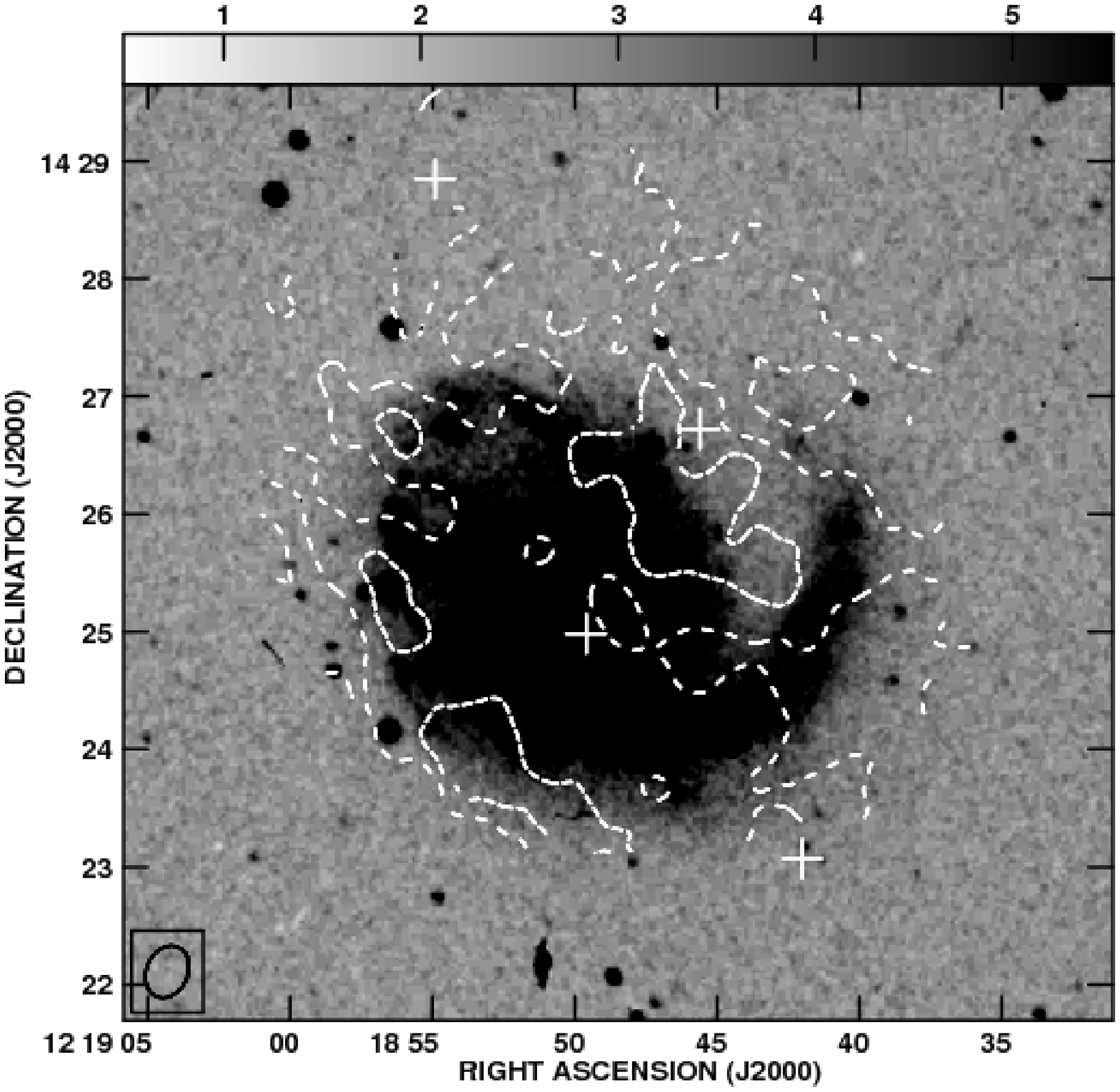}
\includegraphics[width=8cm]{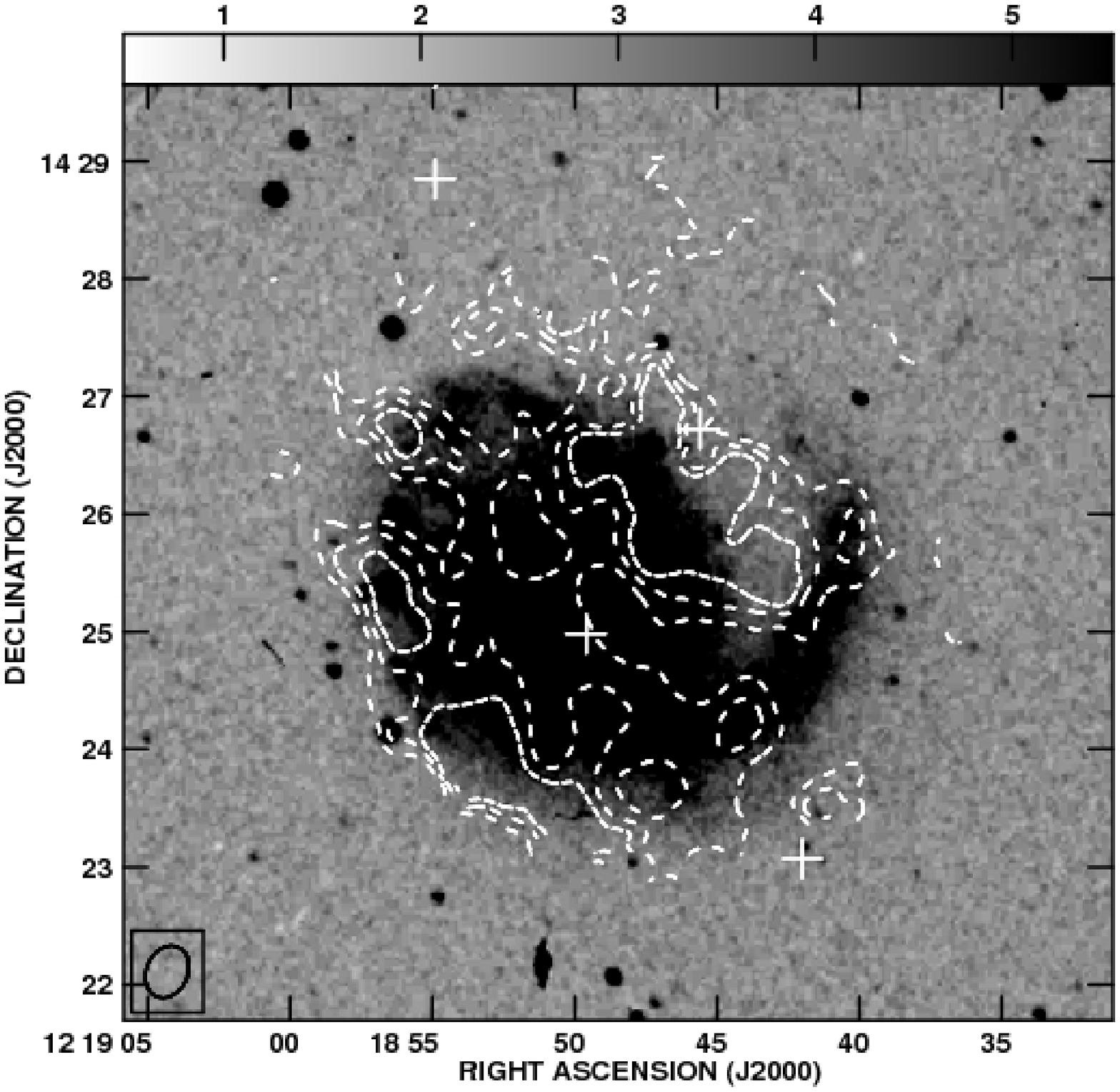}
\caption{ Spectral index maps of NGC 4254. 
The maps were were convolved to a resolution of $27.4''\times21.1''$,
PA$=-24^{\circ}$ and a flux cutoff of  $2\sigma$ was used.
The spectral index contours are superposed on the DSS optical image.
The solid lines outline regions which are blanked or show positive spectral index
whereas the dashed line indicate the negative spectral index distribution.  The
crosses mark the position of the centre of the galaxy and three background
sources which have been removed from the images.
(a) Left panel shows contours for $\alpha=-2,-1$ which is typical
of the gas beyond the optical disk. 
(b) Right panel shows contours for $\alpha=-0.8,-0.4$ which is typical
of the disk gas.  }
\label{fig6}
\end{figure*}

A spectral index map of NGC 4254 was constructed between 240 and 610 MHz
(Fig. \ref{fig6}).  Images at both the frequencies were made
with a maximum baseline of 15 k$\lambda$ and then convolved to 
an angular resolution of $27''\times21''$ at a position angle of $-24^{\circ}$.  
A cutoff of $2\sigma$ was used for both the images. 
While the spectral index across the disk is typically between  $\alpha = -1~$ and $~ -0.4$
with few regions showing $\alpha<-1$,  the 
emission outside the optical disk typically has a spectral index steeper than $-1$ 
(see Fig. \ref{fig6}). 
The spectral index of the envelope is reminiscent of the halo emission in 
edge-on galaxies.
Chyzy et al \nocite{chyzy} (2006,2007) report the detection of a
sharp bright magnetized ridge at 1.43 GHz in the south and present
their image at 4.8 GHz.  They do not detect the southern feature  
at 4.8 GHz which implies a spectral index steeper than $-1.5$ 
when combined with our 240 MHz data.  This is consistent with our data at 610 MHz.  
The spectral index near the centre of the galaxy is comparable to the disk ruling
out the presence of an AGN at the centre. 

From the above discussion, we conclude that there are two components to the radio continuum
emission from NGC 4254, namely the intense disk component which has 
$\alpha \ge -1$ and the extended component which has 
$\alpha \le -1$.  The two components are difficult to distinguish
morphologically.

\section{Discussion}
The radio power at 1.4 GHz of most normal spiral galaxies lie 
between $1.6\times10^{19}$ Watts-Hz$^{-1}$
and  $10^{22}$ Watts-Hz$^{-1}$  (Hummel \nocite{hummel1} 1981) 
and absolute blue magnitudes lie between $-18$ to $-21$ (Hummel \nocite{hummel1} 1981).
NGC 4254 has a power of $1.5 \times 10^{22}$ Watts-Hz$^{-1}$ at 1.4 GHz.
The RC3 (de Vaucouleurs et al. \nocite{vaucouleurs} 1991) 
25 magnitude size of NGC 4254 is $322''\times281''$ and its total blue apparent magnitude
is 10.44 yielding an absolute magnitude of $-21.05$. 
NGC 4254 is thus, both optically and radio bright.
High star formation rates (Schweizer \nocite{schweizer} 1976) of
$\sim 10 M_\odot$yr$^{-1}$ (Soria \& Wong \nocite{soria} 2006) have been reported for NGC 4254.
Cayatte et al. \nocite{cayatte} (1990) suggested that the sharp cutoff in HI in
the south-east could be due to compression by the ICM which
could then explain the bright disk.  
Simulations by Vollmer et al. \nocite{vollmer} (2005) suggest that ram pressure
needs to be invoked to explain the observed HI morphology. 
More recently, Chyzy et al. \nocite{chyzy} (2006,2007) have 
detected a large polarized envelope of emission at 1.43 GHz around the
optical disk and also a bright ridge
of polarised emission in the south which they suggest is due to field compression
by ram pressure. This, they suggest, could be enhanced by the non-disk HI clouds
detected by Phookun et al. \nocite{phookun} (1993) falling back on the disk.   
All the above indicate that ram pressure due to interstellar
medium (ISM)-ICM interaction is active in this system.

On the other hand, ample evidence that this galaxy was involved in a tidal
encounter has been found - the single dominant spiral arm, the presence of
non-disk HI clouds (Phookun et al. \nocite{phookun} 1993) and 
the presence of a large HI tail which extends northwards from NGC 4254 to a distance
of 250 kpc (Haynes et al. \nocite{haynes} 2007).   Haynes et al. \nocite{haynes} (2007)
attribute this long tail to gas stripped from NGC 4254 by 
the process of galaxy harrassment (Moore et al.
\nocite{moore} 1996) as it enters the Virgo cluster at high speed.  
The presence of the large HI cloud VIRGOHI21 (Davis et al. \nocite{davies} 2004) 
which connects spatially and kinematically (Haynes et al. \nocite{haynes}, 
Minchin et al. \nocite{minchin} 2005) to NGC 4254 
resembles the results of the simulations by Bekki et al. \nocite{bekki} (2005).  

The X-ray emission detected from this galaxy (Chyzy et al. \nocite{chyzy1} 2007,
Soria \& Wong \nocite{soria} 2006) is confined to the optical
disk.  This indicates the absence of hot gas
around NGC 4254 which is as expected since it is located on 
the periphery of the Virgo cluster at a projected distance of 1.2 Mpc from M87.
Soria \& Wong \nocite{soria} (2006) suggest that the ultraluminous X-ray (ULX) 
source they detect due south of the optical disk of NGC 4254 has been triggered
by the impact of the infalling HI cloud (Phookun et al. \nocite{phookun} 1993) 
on the disk.  

\subsection{Origin of the extended radio continuum emission}
Edge-on galaxies commonly show the presence of  halo emission extending along the
minor axis which is widely studied at low radio frequencies owing to their
steep spectra. 
We examined the DSS maps of a few low inclination galaxies (e.g. NGC 5236, 
NGC 2997, NGC 6946) 
taken from the atlas by Condon et al. \nocite{condon1} 
(1987) and find that the radio and optical disks have similar extents.  
Hence we find the radio continunum envelope detected around  NGC 4254 
interesting and in this section we try to understand the 
origin of the extended component (Fig. \ref{fig3}).
We believe one of the following scenarios can account for this extended 
radio continuum component.

\subsubsection{Extension of spiral arms}
If we examine Fig \ref{fig3}b closely, it appears that the radio emission extends
further along the spiral arms of the galaxy especially in the north-west and south-east. 
Thus, one possibility is that the spiral arms extend further out in the radio than in the optical.
We examine the plausibility of this as follows.
The typical synchrotron lifetimes of relativistic electrons diffusing
out from supernovae in the galaxy would be $4\times10^8$ years at 240 MHz if the
magnetic field was 3 $\mu$G.  Since we observe
radio continuum emission from the envelope, this suggests that the relativistic electron
population was generated less than $4\times10^8$ years ago.  If this was the case,
then at least the low mass stars  (lifetimes $> 4\times10^8$ years)
from the progenitor stellar population should be surviving even if the
OB associations have disappeared (lifetimes $\sim 10^6$ years) during
the synchrotron lifetime of the electrons.  However no optical emission in the DSS R band
is observed to be coincident with the extended features.    
We also examined the NIR emission in the J, H and K bands from the
2MASS maps which trace emission from low mass stars.  
The NIR morphology of the galaxy is similar to the DSS optical
and no emission coincident with the extended features is detected.  
This indicates that the extended features are unlikely to be
extensions of the galactic spiral arms but are a result of 
some other physical processes.  

\subsubsection{Emission from ram-pressure stripped gas}
The extended emission could be from the magnetized electron gas which has been 
removed from the disk due
to external influences like tidal, ram pressure or turbulent viscous stripping.  This would
lead to the ICM being enriched in magnetic field and matter.
\begin{figure*}
\includegraphics[width=8cm]{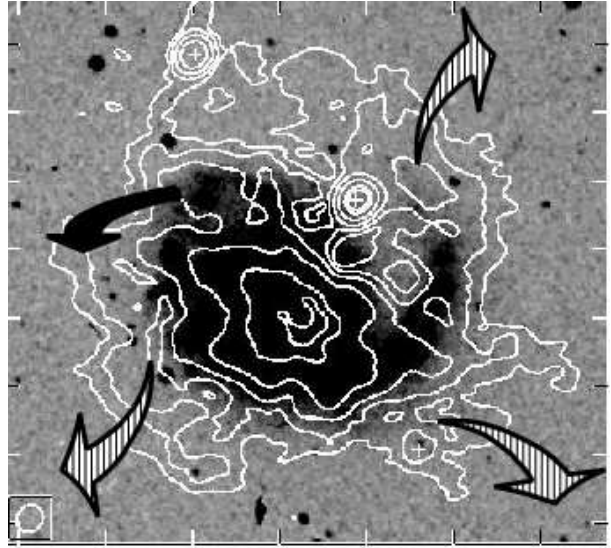}
\caption{Schematic explaining the origin of the extended emission around NGC 4254
observed in radio continuum at 240 MHz.   The 240 MHz map in contours alongwith the optical
DSS map are shown.  The arrows (white with black stripes) indicate the expanding envelope
around NGC 4254 whereas the black arrow shows that the galaxy is moving southwards and
into the page. 
The schematic is not to scale.
The envelope of expanding gas is between us and the galaxy as the galaxy recedes
with a velocity of 1300 kms$^{-1}$ towards the cluster centre.}
\label{fig7}
\end{figure*}

Our multi-frequency radio study of NGC 4254 gives interesting results. 
We detect radio continuum emission from regions surrounding the optical 
disk (see Fig \ref{fig3}).
The spectrum of this emission is steeper ($\alpha \le -1$) compared to the disk emission 
($\alpha \ge -1$).  Moreover,
the extended emission does not form a uniform halo around the optical disk (see Fig. \ref{fig1}). 
While it could be partially attributed to sensitivity and spatial frequency coverage, 
most of this feature  indicates real 
asymmetry in the extended emission.  At 240 MHz, about 15\% of the total emission arises
in this extended component. 
In our HI velocity map, we notice a lagging rotation field in the atomic gas located to the
north of the galaxy.  It is likely that this part is being
stripped and the gas is slowly losing rotation.  

From the above,  we suggest that the extended radio
continuum emission which has a steeper spectral index and the atomic gas in the north of
the galaxy are likely not part of
the disk emission and probably do not lie in the plane of NGC 4254.
This might be gas which has been pushed out of the galaxy by ISM-ICM interaction.
Below we summarize the morphological features that we believe can be
explained by ram pressure caused by the ISM-ICM interaction: 
\begin{itemize}
\item A sharp cutoff is present in both the radio continuum and HI to the south/south-east. 
The galaxy is ploughing into the ICM at an angle of $70^{\circ}$ 
(Vollmer et al. \nocite{vollmer} 2005) which is close to face-on
and the south is the leading edge 
encountering the ICM wind.  The swept-up gas would, thus, give rise to such a smooth boundary. 
\item The enhanced star formation rates across the entire galaxy. 
The wind encounters the entire disk and hence
star formation would have been triggered throughout the disk explaining the
enhanced star formation rates ($\sim 10 M_\odot$yr$^{-1}$, Soria \& Wong \nocite{soria} 2006)
\item Intense radio emission.   This would be a direct result of enhanced star formation
and presence of young massive stars in this galaxy.
\item Extended radio continuum around the optical disk.  This emission is from the
relativistic electron gas which has been influenced by ram pressure and which is 
subsequently diffusing beyond the galaxy as the latter rushes 
towards the cluster centre (see Fig. \ref{fig7}) through the ICM.  The extended emission arises
in gas which is not co-planar with the disk of the galaxy but follows the galaxy.    
The HI Tully-Fisher  relation gives a distance of 16.8 Mpc to NGC 4254 
(Schoniger \& Sofue \nocite{schoniger} 1997) whereas the mean
distance to the Virgo cluster is $20.7\pm 2.4$ Mpc
(Federspiel et al. \nocite{federspiel} 1998).  Thus, NGC 4254 is
closer to us compared to M87 and hence it is reasonable to consider it to be receding and  
falling in towards the centre of the cluster.  The matter giving rise to the extended
radio continuum, then, lies between us and the galaxy. 
\item Steep spectrum of the extended continuum.  The extended emission is similar
to the halo emission observed in edge-on disk galaxies which is also found to
have a steeper spectrum compared to disk emission.  We believe a similar explanation
holds for the emission around NGC 4254.
\item Extended polarized envelope at 1.4 GHz detected around the 
optical disk by Chyzy et al. \nocite{chyzy}\nocite{chyzy1} (2006,2007).  
\item Minor HI deficiency of 0.17 quoted by Cayatte et al. (1994).  
\end{itemize}

Based on the above, we believe that the extended radio continuum emission detected
outside the optical disk is mainly because of a ram pressure event as the galaxy moves
in the ICM. The point we would like to stress is that ram pressure is effective
even though the galaxy is located in the low ICM density
environs in the outskirts of the Virgo cluster.  

Using equipartition and minimum energy arguments, we find that the equipartition 
magnetic field in NGC 4254 is
about 3$\mu$G and the magnetic pressure is $4\times10^{-13}$ ergs-cm$^{-3}$.
The ram pressure acting on the galaxy, which lies about
$3.7^{\circ}$ from the centre of the cluster
was estimated to be $9\times10^{-13}$ ergs-cm$^{-3}$  (Cayatte et al.\nocite{cayatte1} 1994).  
They estimate the gravitational pull for this galaxy to be $16\times10^{-13}$ ergs-cm$^{-3}$. 
Thus, the ram pressure acting on the galaxy leading to gas stripping
is slightly larger than the magnetic pressure holding back the relativistic electrons
from escaping and about half of the gravitational pressure
of the galaxy.  Since this galaxy has undergone a tidal encounter, 
the tidal debris could increase the ICM densities or the tidal
interaction could have reduced the surface density of gas in the outer parts of the
galaxy leading to an enhancement in the ability of ram pressure to 
influence the ISM of NGC 4254.  Similar mechanism has been suggested for
other Virgo galaxies (e.g. Vollmer \nocite{vollmer4} 2003, 
Chung et al. \nocite{chung} 2007) and poor groups 
(e.g. Kantharia et al. \nocite{kantharia} 2005). 
Moreover the high star formation rate of the galaxy would lead to intense
stellar winds and supernova explosions which would further facilitate
the ISM-ICM interaction.
Thus, we believe that our radio observations support the earlier results
of Vollmer et al. \nocite{vollmer} (2005) and Chyzy et al. \nocite{chyzy} (2006) where
both tidal interactions and ram pressure were put forward to explain the observed morphology
of the HI and the radio continuum at 1.4 and 4.8 GHz of NGC 4254.  
We also suggest a possible way to determine whether 
a low inclination galaxy has undergone a ram pressure
event by examining its low radio frequency morphology and the distribution of
the spectral index across it. 

\subsection{Comparison with published simulation results}
Several authors have studied ram pressure stripping in a variety
of environs and with a variety of 
initial conditions in clusters and groups (e.g. Schulz \& Struck \nocite{schulz} 2001,
Vollmer \nocite{vollmer1} 2001, Roediger \& Hensler \nocite{roediger1} 2005, 
Roediger \& Bruggen \nocite{roediger} 2006, Hester \nocite{hester} 2006).
We examine these in light of our new low frequency radio observations of NGC 4254. 
The angle between the disk of NGC 4254 and its orbital plane in Virgo cluster is
$70^{\circ}$ (Vollmer et al. \nocite{vollmer} 2005) and the stripping is close to, but
not entirely, face-on
as the galaxy moves within the cluster.  Roediger \& Bruggen \nocite{roediger} (2006),
in their simulations of ram pressure stripping
in disk galaxies, have shown that for face-on stripping, the stripped gas will expand and
form a tail behind the galaxy.  For a low inclination galaxy,
the stripped gas will form a halo around it over a few hundred 
million years and would be detectable depending on its emissivity. 
Vollmer et al. (2005) performed simulations to explain the observed HI morphology 
(Phookun et al. \nocite{phookun} 1993) of NGC 4254.  They could explain 
most of the morphological
features only if ram pressure is invoked alongwith a tidal encounter 
which would have occurred about 300 Myrs ago.
Their results indicated that ram pressure would only succceed in 
distorting gas with hydrogen column densities $< 10^{20}$ cm$^{-2}$. 
Roediger \& Hensler \nocite{roediger1} (2005) and Roediger \& Bruggen (2006), 
from their simulations on the influence of
the ICM on the cluster members, infer that galaxies undergo 
ram pressure stripping in three stages.  The first stage, they describe as 
the instantaneous stripping phase 
wherein the outer gas disk expands without becoming unbound from the galaxy.  It can
lead to truncation of the gas disk without much mass loss.  
This phase is likely to affect galaxies on the 
outskirts of clusters and in groups and is short-lived,
lasting only for 20 to 200 Myr.  In the intermediate, much longer-lived phase, some of
the expanded gas actually evaporates while some of it falls back on the disk.  
In the final third phase, the galaxy loses mass at a stable rate of 1 M$_\odot$yr$^{-1}$ due
to turbulent viscous stripping (Nulsen 1982). 

Comparing this simulation result with our radio continuum data on NGC 4254, it appears that 
gas stripping in NGC 4254 is in the first phase of evolution.
The extended features which surround the optical disk of NGC 4254 (see Fig. \ref{fig3}) 
are probably due to the relativistic gas which has diffused out from the 
disk but has not yet evaporated.   
The same argument also probably holds for the HI gas.  
The gas is being stripped as seen from the jagged morphology of 
HI that we detect especially in the north and along the vertical edge
in the east.  The smooth boundary to the south is the leading edge and
is caused by the wind blowing across it.  Some of the gas in the north which
shows a lagging rotation field might be due to stripped gas which is slowly losing
rotation.  
We also tried to understand the non-disk HI clouds that Phookun et al. 
\nocite{phookun}(1993)
have detected surrounding the galaxy and not following galaxy rotation.  Phookun et al.
\nocite{phookun} (1993) conclude that these clouds form an 
infalling population.  While the tidal origin put forward by Phookun et al.
\nocite{phookun} (1993) is the most plausible especially since
these form part of the long HI tail which Haynes et al. \nocite{haynes} (2007) attribute
to galaxy harrassment, we examined it in relation to the
three stages of ram pressure stripping inferred by Roediger \& Hensler \nocite{roediger1} (2005)
and explained in brief above.  If we hypothesize that the infalling clouds were
stripped from the galaxy due to ram pressure then the HI gas in
this galaxy might have just entered into the second phase of evolution   
(Roediger \& Hensler \nocite{roediger1} 2005).
In this phase, part of the HI falls back whereas part of the gas is lost from the
system.  Some of the far non-disk HI clouds that Phookun et al. \nocite{phookun}
(1993) detect could also possibly be part of this population.  
Since the galaxy is receding towards south/south-east,
the tail of clouds extending northwards
appears to be in the expected direction.
However, we also note that Roediger \& Bruggen \nocite{roediger} (2006) 
demonstrate in their simulations that the 
direction of motion of the galaxy cannot always be surmised
from the tail of stripped gas.  It may be mentioned, however,
that our explanation of the infalling clouds (Phookun et al. \nocite{phookun} 1993) 
is more speculative and requires further evidence from simulations.  

Since NGC 4254 is located in a relatively low density environment in the outskirts
of the Virgo cluster, the ram pressure stripping event is 
relatively young and was begun 
at most 200 Myrs ago (Roediger \& Hensler \nocite{roediger1} 2005).  
Recall that the simulation by Vollmer et al. \nocite{vollmer} (2005)
indicated that a tidal interaction involving NGC 4254 occurred $\approx 300$ Myrs ago.
These timescales support the hypothesis that the tidal encounter 
boosted the potential of ISM-ICM interaction making it more effective then it would
have otherwise been.  The schematic in Fig. \ref{fig7} summarizes the scenario for
NGC 4254 as discussed above. 

\section{Summary \& Conclusions}
In this paper, we have presented multifrequency radio continuum observations
made with the GMRT of the spiral galaxy NGC 4254 located in the north-west outskirts of the
Virgo cluster.  NGC 4254 is a luminous galaxy with a radio power of
$7\times10^{22}$ Watts-Hz$^{-1}$ at 240 MHz.  
We report the detection of extended, structured, low surface brightness radio continuum 
emission at 240 MHz surrounding the optical disk.  
We note that Chyzy et al. \nocite{chyzy} (2006,2007) have reported detection 
of an extended polarized envelope around the optical disk at 1.4 GHz.   

Combining our data with existing spectral data (Soida et al. \nocite{soida} 1996, Chyzy et al.
\nocite{chyzy1} 2007),
we find that the best fit to the global spectrum is given by a spectral index of 
$-0.848 \pm 0.016$ which is close to the value Soida et al. \nocite{soida} (1996)  quote. 
However from our data at 240, 325 and 610 MHz, we deduce that the spectrum
flattens at the lower frequencies.  A heuristic model which includes electron
propogation effects (Hummel \nocite{hummel} 1991) gives
a better fit to the data.  The injection spectrum has an index $\alpha_0=-0.45\pm0.12$.  

From a detailed spectral index mapping between 240 and 610 MHz, we find that the 
spectral index of the extended emission is steeper with $\alpha \le -1$ compared to
that of the radio emission coincident with the optical disk ($\alpha \ge -1$).   
We hypothesize that the extended emission is not associated
with the optical disk but has been pushed out due to ram pressure of the ICM acting
on the ISM of the galaxy.  The atomic material in the north lags in rotation wrt
rest of the disk and this we believe is part of the gas mass which is being influenced by 
ram pressure. 
NGC 4254 is moving away from us and the stripped material lies between us and
the galaxy as shown in the schematic of Fig. \ref{fig7}. 

From the above, we conclude the following:
\begin{itemize}
\item We believe that we have detected a clear signature of ram pressure stripping in
NGC 4254, a spiral galaxy in the outskirts of the Virgo cluster, in the form of
an extended envelope of emission surrounding the optical disk at 240 MHz.
This emission exhibits a steep spectral index as compared to the disk emission.
We suggest that these might be common signatures of ISM-ICM interaction  
which have not been noticed due to a dearth of such low frequency high 
sensitivity data on low inclination galaxies in cluster/group environments.

\item We suggest that low frequency  ($\le 300$ MHz) high sensitivity survey of
a sample of face-on galaxies in low density groups and clusters will help generate significant
information on the radio frequency signatures of ISM-ICM interactions.  
Signatures such as a steep spectrum and extended emission around low inclination
galaxies can be studied for a larger sample.

\item ISM-ICM interactions are likely to be more common in lower density, lower
velocity dispersion environments than is presently believed.  There are already
several examples (e.g Solanes et al. \nocite{solanes} 2001,
Kantharia et al. \nocite{kantharia} 2005,  Levy et al.
\nocite{levy} 2007) which suggest that ram pressure events are active in
X-ray poor groups and in the outskirts of clusters.  Simulations 
by several groups (e.g. Roediger \& Bruggen 2006, Vollmer et al. 2005, Hester 2006 )
show that this is possible.  Ram pressure in many of these cases
is aided by other physical phenomena such as tidal interaction and supernova
explosions.  The stripped gas would help enhance the ICM densities further.
Systematic low frequency observations of several galaxies in the outskirts of
clusters and in poor groups are required for further understanding.

\end{itemize}

\section{Acknowledgements}
We thank the anonymous referee for a detailed report and
for giving several helpful suggestions on the manuscript. 
We thank the staff of the GMRT that made these observations possible. 
GMRT is run by the National Centre for Radio Astrophysics of the Tata Institute of
Fundamental Research.
This research has made use of the NASA/IPAC Extragalactic Database (NED) and NASA/ IPAC 
Infrared Science Archive
which is operated by the Jet Propulsion Laboratory, California Institute of Technology, 
under contract with the National Aeronautics and Space Administration.
This research has made use of NASA's Astrophysics Data System.
NGK thanks Prof. T. P. Prabhu for useful discussions.

{}


\begin{thebibliography}{}

  \bibitem[2002]{ananth}Ananthakrishnan, S., \& Rao, A. P. 2002, in Multicolour Universe,
International conference on Multi Colour Universe, ed. R. Manchanda \& B. Paul, 233

  \bibitem[2005]{bekki} Bekki, K., Koribalski, B., Kilborn, V., 2005, 2005, MNRAS, 363, L21

  \bibitem[1995]{briggs} Briggs, D., 'High Fidelity Deconvolution of Moderately Resolved
    Sources', PhD thesis, 1995.

  \bibitem[1990]{cayatte} Cayatte, V., van Gorkom, J. H., Balkowski, C., Kotanyi, C., 1990, AJ, 100, 604

  \bibitem[1994]{cayatte1} Cayatte, V., Kotanyi, C., Balkowski, C., 
van Gorkom, J. H., 1994, AJ, 107, 1003

  \bibitem[1943]{chandrasekhar} Chandrasekhar, S., 1943, ApJ., 97, 255

  \bibitem[2006]{chyzy} Chyzy, K.T., Ehle, M., Beck, R., 2006, Astronomische Nachrichten, 327, 501 

  \bibitem[2007]{chyzy1} Chyzy, K.T., Ehle, M., Beck, R., 2007 (arXiv:0708.1533)

  \bibitem[1987]{condon1} Condon, J.J., 1987, ApJS, 65, 485

  \bibitem[1992]{condon} Condon, J.J., 1992, ARA\&A, 30,575

  \bibitem[2007]{chung} Chung, A, van Gorkom, J. H., Kenney, J. D. P., Vollmer, B., 2007,
    ApJ, 659, L115

  \bibitem[2004]{davies} Davies, J., Minchin, R., Sabatini, S., et al., 2004, MNRAS, 349, 922

  \bibitem[1991]{vaucouleurs} de Vaucouleurs, G., de Vaucouleurs, A.,
Corwin, H. G., et al., 1991, Third Reference Catalogue
of Bright Galaxies, (New York; Springer) (RC3)

  \bibitem[1992]{fabbiano} Fabbiano, G., Kim, D.-W., Trinchieri, G., 1992, ApJS, 80, 531

  \bibitem[1998]{federspiel} Federspiel, M., Tammann, G. A., Sandage, A.,
1998, ApJ, 495, 115 

  \bibitem[2007]{gil} Gil de Paz, A., Boissier, S., Madore, B. F., 
et al., to appear in ApJS (astro-ph/0606440)

  \bibitem[2007]{haynes} Haynes, M. P., Giovanelli, R., Kent, B. R., 2007,  ApJ, 665, L19

  \bibitem[2006]{hester} Hester, J. A., 2006, ApJ, 647, 910

  \bibitem[1981]{hummel1} Hummel, E., 1981, A\&A, 93, 93

  \bibitem[1991]{hummel} Hummel, E., 1991, A\&A, 251, 442

  \bibitem[1972]{gunn}Gunn, J. E., \& Gott III, J. R. 1972, ApJ, 176,1

  \bibitem[1986]{hucht} Huchtmeier, W. K., Richter, O.-G., 1986, A\&AS, 64, 111

  \bibitem[2005]{kantharia} Kantharia, N. G., Ananthakrishnan, S., Nityananda, R.,
  Hota, A., 2005, A\&A, 435, 483

  \bibitem[2004]{knapen} Knapen, J. H., Stedman, S., Bramich, D. M., Folkes, S. L., Bradley, T. R., 
   2004, A\&A, 426,1135

  \bibitem[1975]{lecar} Lecar, M., 1975, in
Dynamics of Stellar Systems: Proceedings from IAU Symposium no. 69,
Ed. Avram Hayli,  Dordrecht; Boston: D. Reidel Pub. Co., p.161

  \bibitem[2007]{levy} Levy, L., Rose, J. A., van Gorkom, J. H., Chaboyer, B.,
  2007, AJ, 133, 1104

  \bibitem[2005]{minchin} Minchin, R. F., Davies, J.I., Disney, M. J., et al., 2005, ApJ, 622, L21

  \bibitem[2007]{minchin1} Minchin R. F., Davies, J. I., Disney, M. J. et al., 2007, 
to appear in ApJ (astro-ph/0706.1586)

  \bibitem[1996]{moore}	Moore, B., Katz, N., Lake, G., Dressler, A., Oemler, A., Jr.,
  1996, Nature, 379, 613

  \bibitem[1982]{nulsen}Nulsen, P. E. J., 1982, MNRAS, 198, 1007

  \bibitem[1993]{phookun}Phookun, B., Vogel, Stuart N., Mundy, Lee G., 1993, ApJ, 418, 113

  \bibitem[1991]{pohl}Pohl, M., Schlickeiser, R., Hummel, E., 1991, A\&A 250, 302

  \bibitem[2006]{roediger} Roediger, E., Brüggen, M., 2006, MNRAS, 369, 567

  \bibitem[2005]{roediger1}Roediger, E.,  Hensler, G., 2005, A\&A, 433, 875

  \bibitem[1997]{schoniger} Schoniger, F., Sofue, Y., 1997, A\&A, 323, 14

  \bibitem[2001]{schulz} Schulz, S., Struck, C., 2001, MNRAS, 328, 185	

  \bibitem[2006]{soria}	Soria, R., Wong, D. S., 2006, MNRAS, 372, 1531

  \bibitem[1996]{soida}	Soida, M., Urbanik, M., Beck, R., 1996, A\&A, 312, 409

  \bibitem[2003]{sofue}	Sofue, Y., Koda, J., Nakanishi, H., Hidaka, M.,  2003, PASJ, 55, 75

  \bibitem[1991]{swarup}Swarup, G., Ananthakrishnan, S., Kapahi, V., et al. 1991, 
Current Science, 60, 95 

 \bibitem[1976]{schweizer} Schweizer, F., 1976, ApJS, 31, 313

 \bibitem[2001]{solanes} Solanes, J. M., Manrique, A., Garcia-Gomez, C., et al.
   2001, 548, 97

  \bibitem[1977]{toomre} Toomre, A., 1977, ARA\&A, 15, 437

  \bibitem[1986]{urbanik} Urbanik, M., Klein, U., Graeve, R.,  1986, A\&A, 166, 107

  \bibitem[2004]{urbanik1} Urbanik, M., in Proceedings of 'The Magnetized Plasma
in Galaxy Evolution',  2004, p201, Ed: Chyzy, K. T., et al.

  \bibitem[2001]{vollmer1} Vollmer, B., Cayatte, V., Balkowski, C.,  Duschl, W. J.,  
2001, ApJ, 561, 708 

 \bibitem[2003]{vollmer4}  Vollmer, B., 2003, A\&A, 398, 525

 \bibitem[2004]{vollmer2}  Vollmer, B., Thierbach, M., Wielebinski, R., 2004, A\&A, 418, 1

  \bibitem[2005]{vollmer} Vollmer, B., Huchtmeier, W., van Driel, W.,  2005, A\&A, 439, 921

  \bibitem[2007]{vollmer3} Vollmer, B., Soida, M.,  Beck, R., et al., 2007,  A\&A, 464, L37

  \bibitem[1988]{warmels} Warmels, R. H. 1988, A\&AS, 72, 57

  \bibitem[2007]{wezgoweic} Wezgoweic, M., Urbanik, M., Vollmer, B., et al.,
2007, A\&A, 471, 93

  \bibitem[1988]{wray} Wray, J. D., in 'The Colour Atlas of Galaxies', 1988, Cambridge Univeristy Press.


\end{thebibliography}
\end{document}